\documentclass[aps,prd,superscriptaddress,preprint, showpacs,nofootinbib,floatfix]{revtex4-1}
\pdfoutput=1

\usepackage{amsmath}
\usepackage{amssymb}
\usepackage{mathtools}
\usepackage{graphicx}
\usepackage{relsize}
\usepackage[hidelinks]{hyperref}
\usepackage{wasysym}

\usepackage{color}

\begin{document}


\title{Conformal symmetry and the cosmological constant
 problem~\footnote{Essay written for the Gravity Research Foundation 2018 Awards for Essays on Gravitation.}
}




\rightline{Date: 31.03.2018}

\author{Stefano Lucat}
\email[]{S.Lucat@uu.nl}

\author{Tomislav~Prokopec}
\email[]{T.Prokopec@uu.nl, corresponding author}

\author{Bogumi{\l}a \'{S}wie\.{z}ewska}
\email[]{B.Swiezewska@uu.nl}

\affiliation{Institute for Theoretical Physics, Spinoza Institute
$\&$ EMME$\Phi$, Faculty of Science, Utrecht University,
Postbus 80.195, 3508 TD Utrecht, The Netherlands}

\begin{abstract}

We argue that, when a theory of gravity and matter is  endowed with (classical) conformal symmetry, the fine tuning required to obtain the cosmological constant at its observed value can be significantly reduced. Once tuned, the cosmological constant is stable under a change of the~scale at which it is measured. 
\end{abstract}

\maketitle


\section{'t Hooft's technical naturalness}
\label{tHooft technical naturalness}

In 1979 't Hooft~\cite{tHooft:1979rat} proposed an explanation to why a physical 
parameter may be small. This {\it technical naturalness hypothesis} states that:
\begin{enumerate}
\item[]
{\it A physical parameter $\alpha(\mu)$
 [..] is allowed to be very small only if the replacement $\alpha(\mu)\rightarrow 0$ 
would increase the symmetry of the system. }
\end{enumerate}
't Hooft then observes that the smallness of the parameter is protected in the sense that 
-- due to the enhanced symmetry -- quantum corrections 
will necessarily be proportional to $\alpha(\mu)$ -- and thus will not affect
the smallness of the parameter, explaning the term `technical'. 

In this essay we combine the technical naturalness hypothesis with conformal symmetry 
to argue that a classically conformally invariant
theory of gravity, matter and interactions provides a natural framework in which the cosmological constant 
can be small. In what follows we argue that, once a small cosmological constant 
is generated through radiative breaking of 
conformal symmetry~\cite{Duff:1993wm,Capper:1974ic,Coleman:1973jx}, 
it is protected from growing large ({\it cf.} Ref.~\cite{Koksma:2011cq}).


\section{Conformal symmetry}
\label{Conformal symmetry}

Local Weyl symmetry (in the literature often referred to as conformal symmetry)
is an internal symmetry that -- in addition to diffemorphism invariance -- naturally survives
a breaking of local conformal symmetry which we assume to be realised at very high energies/short distances.
Because space-time torsion changes lengths of parallelly transported tensors, 
torsion tensor is the natural candidate which imbues Weyl symmetry in the gravitational
sector~\cite{Lucat:2016eze}.\footnote{Very much like the longitudinal component 
of the vector field in an Abelian gauge theory, the longitudinal component 
of the torsion trace vector contains the compensating scalar that implements Weyl symmetry to 
Einstein's vacuum equations.} More precisely, only 
the torsion trace part of torsion tensor -- defined by 
 ${\cal T}_\alpha \equiv (2/3)T^\mu_{\;\mu\alpha}$ --
 transforms under Weyl transformations.
That is, if the metric tensor transforms as,
\begin{equation}
 g_{\mu\nu}(x)\rightarrow {\rm e}^{2\theta(x)}g_{\mu\nu}(x)
\,,
\label{Weyl transformation: metric}
\end{equation}
then 
\begin{equation}
 T^\alpha_{\;\mu\nu}\rightarrow  T^\alpha_{\;\mu\nu} +\delta^\alpha_{\;[\mu}\partial_{\nu]}\theta
\;\;\Rightarrow\;\; {\cal T} \rightarrow {\cal T} + d\theta
\,.
\label{Weyl transformation: torsion}
\end{equation}
These transformations then imply that classical gravity in vacuum is conformal.
It is straightforward to extend the symmetry~(\ref{Weyl transformation: metric})--(\ref{Weyl transformation: torsion})
to the matter sector~\cite{Lucat:2016eze}.
The coupling of gravity to matter can then be made conformal by adding a dilaton field 
(whose condensate determines the value of the Newton `constant').
Finally, quantum effects break conformal symmetry~\cite{Capper:1974ic,Lucat:2017wtu}
and in the remainder of the essay we discuss in which way these breakings affect
the observed cosmological constant.


\section{Gravitational hierarchy problem}
\label{Gravitational hierarchy problem}

 
The {\it cosmological constant problem} (CCP) is by far the most severe hierarchy problem of physics,
and up to date no convincing solution has been proposed that is accepted by most physicists. Assuming the observed cosmological constant (CC) is given by dark energy then (in dimensionless units): 
$ (8\pi G_N)\Lambda \sim 10^{-122}$. The CCP can be stated as follows~\cite{Nobbenhuis:2004wn}
(for reviews see also \cite{Weinberg:1988cp,Peebles:2002gy,Rugh:2000ji}): 
\begin{enumerate}
\item[1)] Why is the cosmological constant so small (when measured in natural units)?
\item[2)] Why is it becoming important right now (when we are observing), i.e.\  
why is the energy density in CC so close to the energy density in matter fields, 
$\rho_\Lambda\equiv \Lambda/(8\pi G_N)\sim \rho_m$? 
\item[3)] If CC is diffferent from zero, what sets its magnitude and what stabilizes it against running with the energy scale? 
\end{enumerate}

To elaborate on {\it Problem 1}, note that quantum vacuum fluctuations contribute 
to $\rho_\Lambda$ as $\sim k_{\mathrm{UV}}^4$, where $k_{\mathrm{UV}}$ is an ultraviolet momentum cutoff scale.
Given that the natural cutoff of quantum gravity is the Planck scale (the scale at which gravity 
becomes strongly interacting), $m_{\rm P} = 1/\sqrt{G_N}$, the first problem 
can be rephrased as: {\it Why is $\Lambda/m_{\rm P}^2\ll 1$?} In other words, why 
vacuum fluctuations do not (significantly) contribute to $\Lambda$?
As regards {\it Problem 3}, we note that, if one can identify the symmetry which 
is realised when CC vanishes, then this symmetry protects CC from running fast with scale.

\section{Conformal symmetry and cosmological constant}
\label{Conformal symmetry  and cosmological constant}

 If a theory of gravity, matter and interactions is classically conformal, still quantum effects can violate the classical symmetry. The couplings constants can start running in such a way that the potential develops a new minimum, away from the origin of the field space thus introducing an energy scale and breaking conformal symmetry~\cite{Coleman:1973jx}. If the couplings are small at some fiducial large energy scale $\mu_*$, their running will typically be slow, allowing for a large hierarchy between the UV scale and the scale of symmetry breaking. The latter can be estimated as the scale at which a given coupling turns negative, allowing for a minimum to form. From the perturbative treatment of the running we obtain an estimate on that scale,
$\mu\sim\mu_*\exp(-1/\lambda_*)$, where $\lambda_*$ denotes the relevant coupling at the 
scale $\mu_*$ and we have dropped factors of ${\cal O}(1)$ in the exponent. 
Assuming that the vacuum expectation value of the scalar field is of the order of the scale $\mu$ we obtain a rough estimate
\begin{equation}
\rho_{\rm \Lambda}\sim - v^4\sim -\mu^4 \sim -\mu_*^4\exp(-\frac{4}{\lambda_*}).
\end{equation}
In light of the above and with the right choice of the couplings at $\mu_*$
($\lambda_*\sim 10^{-2}$), 
one could, in principle, get a cosmological constant as small as the observed one (though negative). 
In practice, however, Nature has chosen to break conformal symmetry in the matter sector 
at the electroweak scale, $\mu\sim 10^2~{\rm GeV}$, 
at which the Higgs field acquires an expectation value of $\langle h\rangle\equiv v \simeq 246~{\rm GeV}$, 
which is responsible for the mass generation of all standard model particles 
(except perhaps of the neutrinos).\footnote{In the case of conformally symmetric theory the BEH mechanism of the standard model is replaced by the Coleman--Weinberg mechanism~\cite{Coleman:1973jx}.}
This then sets the natural energy density scale for the cosmological constant, 
\begin{equation}
\rho_{\rm \Lambda}^{\rm EW}\sim - v^4\sim -10^8~{\rm GeV^4}.
\label{rho matter}
\end{equation}
The contribution~(\ref{rho matter}) must be negative, since 
in the absence of Higgs condensate $\rho_{\rm \Lambda}^{\rm EW}$ must vanish.\footnote{Another negative contribution is generated 
by the chiral condensate of mesons generated as the chiral symmetry of QCD gets broken by the chiral anomaly but 
-- when compared with~(\ref{rho matter}) -- 
that contribution can be neglected since it is of the order $ - 10^{-4}~{\rm GeV}$.}

 It seems that we have a {\it no-go} theorem: The contribution from matter field condensates is necessarily 
large and negative, while the observed cosmological constant is small and positive. In order to 
overcome this impasse, we ought to dig deeper into the model and understand how gravity 
contributes to the cosmological constant. To get a clearer picture, let us consider the following simple 
conformal model of gravity consisting of the metric field $g_{\mu\nu}$, 
the torsion  field $T^\alpha_{\;\rho\sigma}$,
the dilaton $\phi$ and matter fields $\psi_i$. The action is given by~\cite{Lucat:2016eze}, 
\begin{equation}
S[\phi,g_{\mu\nu},T^\alpha_{\;\rho\sigma},\psi_i] 
= \int \sqrt{-g}d^4x\left\{\frac{\alpha}{2}\phi^2 R + \frac{\beta}2 R^2 
 -\frac{1}{2}g^{\mu\nu}\nabla_\mu\phi\nabla_\nu\phi -\frac{\lambda}{4}\phi^4\right\} +S_m[\psi_i,g_{\mu\nu},T^\alpha_{\;\rho\sigma}]
\,,
\label{conformal action}
\end{equation}
where $R=R[g_{\mu\nu}, T^\alpha_{\;\rho\sigma}]$ is the Ricci curvature scalar, 
$g={\rm det}[g_{\mu\nu}]$,  $\nabla_\mu=\partial_\mu + {\cal T}_\mu$ is the conformal covariant derivative
(${\cal T}_\alpha \equiv (2/3)T^\mu_{\;\mu\alpha}$) and $S_m$ denotes the matter action.

 This action can be discerning for inflation~\cite{Sandro:2018}, but it can be also used to get an insight 
on how the gravitational sector contributes to the cosmological constant. 
The action~(\ref{conformal action}) contains three scalars: 
$\phi$, $R$ (scalaron) and the longitudinal component of torsion trace, 
\begin{equation}
   {\cal T}_\mu^L=\partial_\mu\theta(x)
\,. 
\label{torsion trace: L}
\end{equation}
In the absence of scalar condensates, 
the theory~(\ref{conformal action})
is at its conformal fixed point, and the vacuum energy must vanish. 
In order to understand what happens away from the conformal point, it is instructive to
replace $R$ in~(\ref{conformal action}) by a scalar field $\Phi$. 
This can be done by exacting $R\rightarrow \Phi$ and 
by adding a lagrange multiplier term to the lagrangian in~(\ref{conformal action}), 
$\Delta{\cal L} = \frac12\omega^2(R-\Phi)$. Varying the resulting action 
with respect to $\Phi$ then gives, 
$\Phi=-[\alpha\phi^2-\omega^2]/(2\beta)$. Inserting this into~(\ref{conformal action}) yields 
an (on-shell) equivalent action for the gravitational sector,  
\begin{equation}
S_g[\phi,g_{\mu\nu},T^\alpha_{\;\rho\sigma}] 
= \int \sqrt{-g}d^4x\left\{\frac{\omega^2}{2}R - \frac{1}{2}g^{\mu\nu}\nabla_\mu\phi\nabla_\nu\phi 
-\frac{1}{8\beta}\left(\alpha\phi^2-\omega^2\right)^2
-\frac{\lambda}{4}\phi^4\right\} 
\,.
\label{conformal action:2}
\end{equation}
This (Einstein frame) action is (classically) conformal only if the Lagrange multiplier field transforms as, 
$\omega\rightarrow \Omega^{-1}\omega$ and it reduces to the usual general relativity coupled to 
a real scalar in a gauge 
in which $\omega$ is gauge fixed to the (reduced) Planck mass, 
\begin{equation}
\omega=M_{\rm P}\qquad ({\rm gauge\;\, fixing})
\,, 
\label{gauge fixing}
\end{equation}
with $M_{\rm P}\equiv 1/\sqrt{8\pi G_N}$
(any choice for $\omega$ is in principle allowed). 
Since $M_{\rm P}$ is the only scale in the problem,
it has no absolute meaning, {\it i.e.} conformal symmetry of~(\ref{conformal action:2}) 
teaches us that the choice~(\ref{gauge fixing}) is {\it physically equivalent} to any other non-vanishing 
(local) scale $\omega^\prime(x)=\Omega^{-1}(x)M_{\rm P}$. 
The two remaining scalars, $\theta$ and $\phi$, are the physical (scalar) degrees of freedom of the theory.
Since $\theta$ stems from Weyl symmetry, $\theta$ retains a flat direction, {\it i.e.} it exhibits a (global) 
shift symmetry, $\theta(x)\rightarrow \theta(x)+\theta_0$ and thus cannot contribute to the cosmological 
constant. On the other hand, $\phi$ exhibits a nontrivial potential. A simple calculation shows that,
when $\beta/\alpha>0$, $\phi$ is tachyonic and condenses to, 
$\phi_0^2=\alpha\omega^2/(\alpha^2+2\lambda\beta)$, at which the mass and potential energy
are given by,
\begin{equation}
m(\phi_0)^2 = \frac{\alpha}{\beta}\omega^2
\,,\qquad 
  V(\phi_0) = \frac{\lambda}{4(\alpha^2+2\lambda\beta)}\omega^4
\,.
\label{mass and potential energy}
\end{equation}
Let us pause to try to understand the result~(\ref{mass and potential energy}).
As a consequence of conformal symmetry breaking, the gravitational sector produces 
a positive cosmological constant
whose size in natural (dimensionless) units is given by, 
\begin{equation}
   \frac{ V(\phi_0)}{\omega^4} = \frac{\lambda}{4(\alpha^2+2\lambda\beta)}
\,.
\label{CC size}
\end{equation}
For this to compensate the negative cosmological constant generated in the matter sector, 
one ought to fine tune~(\ref{CC size}) to be $\sim 10^{-65}$, such that when~(\ref{CC size}) 
 is added to the matter contribution~(\ref{rho matter}), one obtains the observed
cosmological constant,  $\Lambda/\omega^4\sim 10^{-122}$. 
We emphasize that, once the cosmological constant is tuned to the observed value, 
't Hooft's technical naturalness ensures that quantum corrections 
(both from gravitational and matter fields) 
will not affect it. This can be made more precise as follows. Let us assume that 
$V_{\rm eff}\sim 10^{-122}\omega^4$ represents the total contribution 
to the cosmological constant. Then, 
the RG improved $V_{\rm eff}$ must obey 
the Callan-Symanzik equation,
\begin{equation}
\mu\frac{d}{d\mu}V_{\rm eff}(\phi,{\rm all\; other\;fields})
   =  \mu\frac{d}{d\mu}V_{\rm eff}(0,{\rm all\; other\;fields\rightarrow fixed \; point})
\,,
\label{CS equation}
\end{equation}
 where the second term constitutes $V_{\rm eff}$ with all fields set to their respective conformal
fixed point (at which all scalars vanish), which 
vanishes for the conformal theory under study.\footnote{If the potential is nonzero at the origin 
of the field space, one has to cancel the zero-point energy order by order to obtain 
a homogenous RG equation for the effective potential~\cite{Ford:1992mv}.}
Eq.~(\ref{CS equation}) tells us then that $V_{\rm eff}$ (and therefore also $\Lambda$)
does not change if the scale $\mu$ changes. This means that, while
the precise value of the fields and couplings can depend on $\mu$, the value of the effective potential
at its minimum cannot.\footnote{In practical applications $V_{\rm eff}$ is always approximated 
by its truncated version at a finite order in loop expansion. Such a truncated effective potential contains 
some residual dependence on $\mu$~\cite{Chataignier:2018aud}, 
which is however suppressed by a suitable power of $\hbar$.}

To conclude, let us recall that the classical conformal symmetry alleviates the hierarchy problem 
associated with the mass of the Higgs boson present in the standard model~\cite{Bardeen:1995kv}. 
In this way imposing conformal symmetry on physical theories can 
elucidate the most notorious hierarchy problems in physics 
-- the cosmological constant problem and the gauge hierarchy problem.


\begin{acknowledgments}
The authors acknowledge supprot from the D-ITP consortium, 
a program of the NWO that is funded by the Dutch Ministry of 
Education, Culture and Science (OCW). This work is part of the research programme of the Foundation for Fundamental Research on Matter (FOM), which is part of the Netherlands Organisation for Scientific Research (NWO). S.L. acknowledges financial support from an NWO-Graduate Program grant.
B.{\'S}. acknowledges support from the National Science Centre, Poland, the HARMONIA project under contract UMO-2015/18/M/ST2/00518 (2016-2019). 
\end{acknowledgments}


\end{document}